# Dynamical Responses of a Valveless Fluid Loop Excited by the Impact of a Compression Actuator

## Chi-Chung Wang* and Tian-Shiang Yang**

## ABSTRACT

Valveless pumping assists in fluid transport in various organisms and engineering systems. In a previous work, to study the actuator impact effects on valveless pumping, we constructed a piecewise-linear lumped-parameter model for a closed-loop system, which consists of two distensible fluid reservoirs connected by two rigid tubes. The preliminary asymptotic and numerical results of that work indicated that the system dynamics is rather rich and complex, and strongly depends upon the driving frequency of the actuator (which periodically compresses one of the distensible reservoirs) and other system parameters. Here, a more comprehensive numerical study on the aforementioned model is carried out, so as to clarify how many different types of system responses can be excited, and to locate the parameter boundaries within which each type exists. Moreover, by examining the driving-frequency dependences of a number of characteristic phases (such as the phase of the compression cycle at which the actuator collides with or separates from the distensible reservoir it acts upon), the interrelations between different types of system responses are identified.

## INTRODUCTION

In various organisms and engineering systems, the working fluid transport more or less is assisted by valveless pumping. For example, even when the heart's valves fail, blood circulation in the cardiovascular system still is maintained to some extent (see Liebau, 1954, and Moser, 1998). Various other biomedical systems involving valveless pumping also are discussed in Moser (1998). Meanwhile, in microfluidics, valveless "impedance pumps" have been fabricated by Rinderknecht et al. (2005), and it is expected that they would be particularly suitable for handling sensitive biofluids. One way to achieve valveless pumping is to compress a pliant tube periodically at an asymmetric site from the tube's interfaces to different tubing or reservoirs, so that pressure waves are excited and propagate on the tube. Due to impedance mismatch at the tube ends, the waves would partly reflect, and hence build up a mean pressure head that drives the fluid flow in the tube (see Hickerson et al., 2005, Hickerson and Gharib, 2006, and Wen and Chang, 2009).

\* *PhD Candidate, Department of Mechanical Engineering, National Cheng Kung University, Tainan, Taiwan 701.*

\*\* *Professor, Department of Mechanical Engineering, National Cheng Kung University, Tainan, Taiwan 701.*

Despite the existence of a sizable research literature on valveless pumping (for a brief review, see Yang and Wang, 2010, and references cited therein), a full understanding of its mechanisms has not yet been obtained. To a certain extent, this is because that valveless systems may differ in configuration and operation conditions (e.g., the compression frequency; see Hickerson et al., 2005, and Wen and Chang, 2009), and therefore rely upon one or more different mechanisms to pump fluids. It is intuitively clear, however, that for a valveless pumping system to produce any mean flow, a necessary condition (but not necessarily a sufficient one; see Yang and Wang, 2010) is to drive (i.e., compress) the system at an asymmetric site with respect to its configuration. It is also arguable that appropriate theoretical modeling is as important as extensive experimental studies in furnishing a more complete understanding of valveless pumping mechanisms.

As pointed by Yang and Wang (2010), in existing mathematical models of valveless pumping, the interaction between the compression actuator and the pliant part of the system in question has often been modeled in rather oversimplified ways. Specifically, in some models the instantaneous deformation of the pliant part of the system is prescribed in a manner not inconsistent (but not absolutely consistent either, as demonstrated by Yang and Wang, 2010) with the actuator motion (see, e.g., Ottesen, 2003, and Huang et al., 2010), while in some other models the temporal variation of the external pressure at the compression site is specified (e.g., Timmermann and Ottesan, 2009). Meanwhile, in the wave-pulse model of Hickerson and Gharib (2006), a pair of pressure waves of a prescribed waveform are sent out from the compression site at the driving frequency. While the actuator and the pliant part of the system may indeed interact in a rather simple way under suitable conditions, it was demonstrated by Yang and Wang (2010) that there are cases in which the collision of the compression actuator and the system's pliant part is the only cause that produces a mean flow in the system.

As the present work extends that of Yang and Wang (2010), let us briefly outline here their model system and major findings. The model system they considered consisted of two distensible reservoirs connected by two rigid tubes, with one of the reservoirs (hereafter referred to as the $V_0$ reservoir) compressed periodically by an actuator at a prescribed frequency (see Fig. 1). It was assumed that the actuator motion sets the upper bound of the instantaneous volume of the $V_0$ reservoir. Moreover, during the whole compression cycle, that upper bound is smaller than the free (i.e., unstressed) volume of the $V_0$ reservoir, and so the nominal duty cycle of the actuator is unity. To suppress wave propagation effects, the driving frequency was assumed to be sufficiently low, and a lumped-parameter model (with constant coefficients) accounting for mass and momentum balance in the system was constructed. Based upon such a model, it was shown that the $V_0$ reservoir may separate from the compression actuator during some part of the compression cycle if the driving frequency of the actuator exceeds a certain threshold value. The actual duty cycle (i.e., the fraction of the compression cycle during which the actuator actually compresses the system's pliant part) would then become less than unity. In fact, only when that happens would a nonzero mean flow be produced in the fluid loop by the synergetic interaction between configurational asymmetry and the nonlinear effects associated with actuator impact. Moreover, the preliminary asymptotic and numerical results of Yang and Wang (2010) indicated that the system dynamics is rather rich and complex, and strongly depends upon the driving frequency of the actuator and other system parameters. (More technical details of their model and findings will be discussed in the next section.)

In this paper, we shall present the results of a more comprehensive numerical study on the model of Yang and Wang (2010). The purposes are to clarify how many different types of system responses can be excited, and to locate the parameter boundaries within which each type exists. Such a study is important because a good knowledge about the qualitative and quantitative behaviors of the system responses is important for designing an efficient valveless pump under certain prescribed specifications. Moreover, by examining the driving-frequency dependences of a number of characteristic phases (such as the particular phase of the compression cycle at which the actuator collides with or separates from the $V_0$ reservoir), the interrelations between different types of system responses will be identified. Such interrelations clearly are of interest to researchers of nonlinear system dynamics.

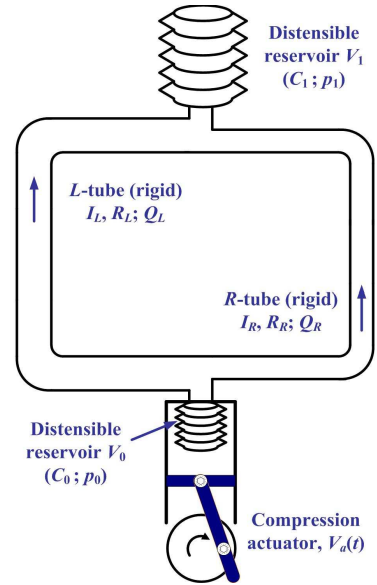

Fig. 1. Schematic of the model system studied here.

## MATHEMATICAL MODEL

As in Yang and Wang (2010), here we consider a closed-loop valveless pumping system that consists of two distensible fluid reservoirs (such as bellows) connected by two rigid tubes; see Fig. 1. The system is completely filled with an incompressible liquid (say, water) of density $\rho$ and viscosity $\mu$. The rigid tube on the left-hand side of the system (hereafter referred to as the $L$-tube) has radius $a_L$ and length $l_L$, while that on the right-hand side (the $R$-tube) has radius $a_R$ and length $l_R$. The $V_0$ reservoir interacts with an external compression actuator, and has a temporally varying volume $V_0(t)$. Similarly, the instantaneous volume of the other reservoir (the $V_1$ reservoir) is denoted by $V_1(t)$.

### Constitutive equations

It is assumed here that the physical presence of the actuator sets the maximum admissible volume of the $V_0$ reservoir at all times. Here we denote such a maximum admissible volume by $V_a(t)$. Accordingly, when $V_0(t) < V_a(t)$, the $V_0$ reservoir will not be in contact with the actuator, and so its volume would vary freely (but in a way that conserves mass and satisfies momentum balance; to be discussed below). For convenience, we shall then say that the system is in the "free mode." On the other hand, whenever the $V_0$ reservoir is in contact with the actuator—and the system is said to be in the "contact mode"—we shall simply have $V_0(t) = V_a(t)$. As a particular example, it is specified here that $V_a(t)$ varies sinusoidally, with an angular frequency $\omega$, between $V_{min}$ ($>0$) and $V_{max}$ ($> V_{min}$):

$$V_a(t) = \frac{1}{2}(V_{max} + V_{min}) + \frac{1}{2}(V_{max} - V_{min})\cos\omega t. \quad (1)$$

Moreover, the instantaneous reservoir volumes $V_{0,1}(t)$ are assumed to vary linearly with the pressure difference across the reservoir walls. Specifically, with $p_0(t)$ being the fluid pressure inside the $V_0$ reservoir, and $p_e(t)$ its external pressure (exerted by the actuator), one has

$$V_0(t) = V_{0f} + C_0 \cdot \{p_0(t) - p_e(t)\}, \quad (2)$$

where $V_{0f}$ is the free (i.e., unstressed) volume of the $V_0$ reservoir, and $C_0$ is a (constant) compliance coefficient. It is assumed here that the actuator can only exert a positive pressure on the $V_0$ reservoir, so that $p_e > 0$ when the system is in the contact mode, and $p_e = 0$ in the free mode. Meanwhile, since the $V_1$ reservoir has an internal pressure $p_1(t)$ but no external pressure, its constitutive relation is written as

$$V_1(t) = V_{1f} + C_1 \cdot p_1(t), \quad (3)$$

with $C_1$ being a constant compliance coefficient.

**Mass and momentum balances**

Now, denoting the volumetric flowrates (from the $V_0$ reservoir to the $V_1$ reservoir) in the $L$- and $R$-tubes by $Q_L(t)$ and $Q_R(t)$, respectively, mass conservation requires that

$$\dot{V}_0 = -Q_L - Q_R, \quad \dot{V}_1 = Q_L + Q_R, \quad (4a, 4b)$$

where the overdots represent derivatives of the state variables with respect to time $t$. Meanwhile, to suppress all nonlinearities except that arising from the interactions between the actuator and the $V_0$ reservoir, the momentum equations for the $L$- and $R$-tubes are given by

$$I_L\dot{Q}_L + R_LQ_L = p_0 - p_1, \quad (5a)$$

$$I_R\dot{Q}_R + R_RQ_R = p_0 - p_1, \quad (5b)$$

where $R_{L,R}$ and $I_{L,R}$ are the (constant) resistance and inertia coefficients, respectively. It can be readily deduced from Eq. (5a) that, if the $L$-tube stands alone (i.e., not connected to the system), $Q_L(t)$ would vary on the intrinsic time scale of $T_L = I_L / R_L$. Similarly, by Eq. (5b), the intrinsic time scale for the variation of $Q_R(t)$ when the $R$-tube stands alone is $T_R = I_R / R_R$.

**Initial conditions**

As in Yang and Wang (2010), here we shall consider the particular case that both the $V_0$ and $V_1$ reservoirs are filled to their free volumes before the compression actuator is installed. The total fluid volume in the two reservoirs, $V_{tot} = V_0(t) + V_1(t) = V_{0f} + V_{1f}$, therefore is invariant with time as can be readily deduced from Eqs. (4a) and (4b). Suppose then that an actuator is installed and compresses the $V_0$ reservoir to a certain extent before it starts moving. In terms of the "pre-compression ratio" $\gamma = (V_{0f} - V_{max})/V_{0f}$ ($>0$), the initial volumes of the $V_0$ and $V_1$ reservoirs therefore are given by $V_0(0) = V_{0f}(1-\gamma)$ and $V_1(0) = V_{1f} + \gamma V_{0f}$, respectively. Also, Eq. (1) can be rewritten as

$$V_a(t) = V_{0f}(1-\gamma) - \Delta V_a(1 - \cos\omega t), \quad (6)$$

where $\Delta V_a = (V_{max} - V_{min})/2$. Meanwhile, one has $Q_L(0) = 0 = Q_R(0)$, and it can be calculated from Eqs. (2) and (3) that initially the uniform internal pressure of the system is $p_0(0) = \gamma V_{0f} / C_1 = p_1(0)$, and the external pressure exerted by the actuator on the $V_0$ reservoir is $p_e(0) = \gamma V_{0f} / C_{01}$, with the "characteristic compliance" of the system $C_{01}$ given by $1/C_{01} = 1/C_0 + 1/C_1$.

With the initial conditions of the system specified above, Eqs. (2)–(6) can be integrated numerically (say, by use of the fourth-order Runge–Kutta method). However, as shown in Yang and Wang (2010), for such a pre-compressed system if the driving frequency $\omega$ is below a well defined threshold value $\omega_{th}$, the $V_0$ reservoir would never separate from the compression actuator, and the perfectly linear governing equations would produce zero mean flowrate in the fluid loop. As a matter of fact, since the purpose of this work is to examine the effects of actuator impact upon valveless pumping, the pre-compressed system is deliberately chosen, such that a nonzero mean flow in the system can only be produced by the interactions between the actuator and the $V_0$ reservoir. Figures 2(a) and 2(b) summarize the possible interactions between the actuator and the $V_0$ reservoir. Since such interactions have been discussed in detail in Yang and Wang (2010), we shall only briefly highlight their physical meanings below.

**Mode decision rules**

First, suppose that at the present time step the system is in the contact mode. As explained above, one will then have $V_0(t) = V_a(t)$, and it follows from Eq. (4a) that $Q_R(t) = -\dot{V}_a(t) - Q_L(t)$. Moreover, the pressure difference $p_0(t) - p_1(t)$ can be eliminated from Eqs. (5a) and (5b), yielding $I_L \dot{Q}_L + R_L Q_L = I_R \dot{Q}_R + R_R Q_R = -I_R(\ddot{V}_a + \dot{Q}_L) - R_R(\dot{V}_a + Q_L)$. With the particular choice of $V_a(t)$ given by Eq. (6), this result can be written as

$$(I_L + I_R)\dot{Q}_L + (R_L + R_R)Q_L$$
$$= \Delta V_a(\omega R_R \sin \omega t + \omega^2 I_R \cos \omega t), \quad (7)$$

from which the value of $Q_L$ at the next time step can be readily calculated by use of the fourth-order Runge–Kutta method. (Incidentally, Eq. (7) also shows that in the contact mode the characteristic time scale for the flowrate variation in the closed fluid loop is $T_{LR} = (I_L + I_R)/(R_L + R_R)$.) The values of other state variables at the next time step then can be calculated in turn following the procedures summarized in Fig. 2(a). If a negative value of the external pressure $p_e$ is obtained at the next time step, it is interpreted that the system switches to the free mode (otherwise the system stays in the contact mode at the next time step). Of course, when a negative value of $p_e$ arises, one would also need to determine the precise "separation time" ($T_s$, when $p_e = 0$) by interpolation; see Fig. 2(a).

Now, if the system is in the free mode at the present time step, then the external pressure $p_e(t) = 0$. Accordingly, it is deduced from Eqs. (2) and (3) that $p_0(t) - p_1(t) = -[V_{0f} - V_0(t)]/C_{01}$, which can be substituted into Eqs. (5a) and (5b) to yield

$$I_L \dot{Q}_L + R_L Q_L = -[V_{0f} - V_0(t)]/C_{01}, \quad (8a)$$

$$I_R \dot{Q}_R + R_R Q_R = -[V_{0f} - V_0(t)]/C_{01}. \quad (8b)$$

Note that the values of $V_0(t)$, $Q_L(t)$ and $Q_R(t)$ at the next time step can be calculated by integrating Eqs. (4a), (8a), and (8b). The values of other state variables at the next time step then can be calculated in turn following the procedures summarized in Fig. 2(b).

It should also be noted that if the system currently is in the free mode, there are more decisions to be made regarding the system state at the next time step. Specifically, in the free mode, the volume of the $V_0$ reservoir may reduce to zero or increase to meet its instantaneous upper bound $V_a(t)$ set by the actuator motion; the latter is interpreted as a collision of the $V_0$ reservoir with the actuator. In the computations, if $V_0(t)$ becomes negative in the next time step, we shall determine first the true "depletion time" $T_0$, with $V_0(T_0) = 0$, by interpolation. Moreover, it is assumed that as soon as the $V_0$ reservoir is completely depleted, the fluid loop is blocked, so that $Q_L(T_0) = 0 = Q_R(T_0)$ and, of course, $V_1(T_0) = V_{0f} + V_{1f}$. For $t = T_0^+$, the system is restarted with the state variable values given above, and then continues its journey in the free mode (see Fig. 2(b)).

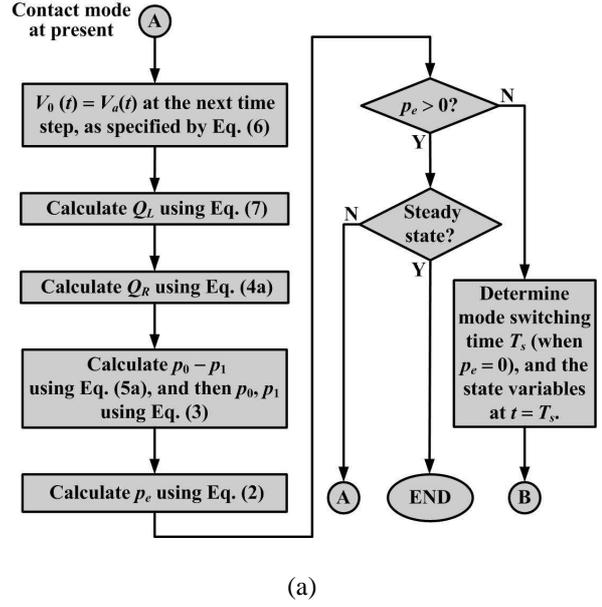

(a)

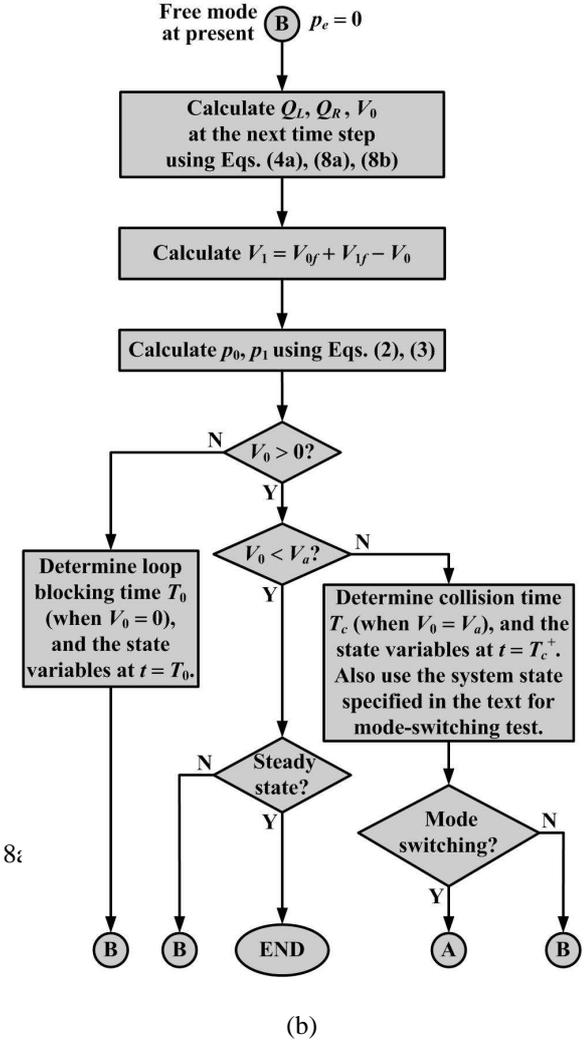

(b)

Fig. 2. Numerical solution procedures and mode decision rules.

Consider next the possibility of collision between the $V_0$ reservoir and the actuator. Again, in the computations it may occur that at the next time step $V_0(t) > V_a(t)$, and interpolation is needed to determine the true collision time $T_c$, with $V_0(T_c) = V_a(T_c)$. Note also that the mechanical impact associated with the $V_0$ reservoir–actuator collision at $t = T_c$ would cause the flowrates in the rigid tubes to vary discontinuously. Specifically, as calculated by Yang and Wang (2010), one would have

$$Q_L(T_c^+) = \frac{I_L Q_L(T_c^-) - I_R[Q_R(T_c^-) + \dot{V}_a(T_c)]}{I_L + I_R}, \quad (9a)$$

$$Q_R(T_c^+) = \frac{I_R Q_R(T_c^-) - I_L[Q_L(T_c^-) + \dot{V}_a(T_c)]}{I_L + I_R}. \quad (9b)$$

Another important calculation here is the system mode following the collision. Specifically, after the collision, the system may switch to the contact mode, or stay in the free mode. To decide what would happen, in the computations we assume first that the system stays in the free mode at $t = T_c^+$, and tentatively continue the computation with $p_c = 0$. If it is calculated at the tentative time step that $V_0(t) < V_a(t)$, we shall take that the system really would stay free; otherwise the system would switch to the contact mode (see Fig. 2(b)).

Clearly, the overall mathematical model involves rather intricate decision rules (summarized in Figs. 2(a) and 2(b)), and its solution generally has to be obtained numerically. However, for a set of realistic sample parameter values (the same as that specified in the next section), Yang and Wang (2010) identified a small dimensionless compliance parameter, and therefore were able to carry out a series of asymptotic calculations to obtain accurate estimates for some critical driving frequencies across which the system responses undergo qualitative changes. Furthermore, numerical solutions of the model for three representative cases (with $V_{min}/V_{max} = 0.6$, 0.33, and 0.15) were carefully discussed to elucidate the complex dynamics of the model system, and to illustrate the frequency dependence of the mean flowrate and mean pressures in the system. Their numerical results indicated that there may exist steady periodic system responses having a period that is a multiple of the compression cycle period at higher driving frequencies. In fact, under suitable conditions, seemingly chaotic system responses also arose in their computations. However, the transition routes of the various types of system responses were not fully discussed by Yang and Wang (2010). Moreover, for steady periodic system responses repeating themselves in one compression cycle, Yang and Wang (2010) had not determined the precise boundaries between different types of system responses. These, then, are the main purposes of the present work.

## SAMPLE PARAMETER VALUES

In our computations, a sample system is envisioned which has realistic dimensions and material properties. Specifically, it is supposed that the $V_0$ and $V_1$ reservoirs are made of bellows. The free (i.e., unstressed) radius and height of the $V_0$ reservoir are $r_0 = 25$ mm and $h_0 = 33.3$ mm, respectively, while the corresponding dimensions of the $V_1$ reservoir are $r_1 = 50$ mm and $h_1 = 66.7$ mm. The free volumes of the two reservoirs therefore are $V_{0f} = 65.4$ ml and $V_{1f} = 523.6$ ml, respectively. Suppose also that the two fluid reservoirs (bellows) are reinforced by elastic springs so that they have the same effective spring constant of $k_0 = k_1 = 15.74$ N/mm. Accordingly, the compliance coefficients of the two reservoirs are estimated from the simple formulae $C_{0,1} = \pi^2 r_{0,1}^4 / k_{0,1}$, yielding $C_0 = 2.45 \times 10^{-4}$ ml/Pa and $C_1 = 3.93 \times 10^{-3}$ ml/Pa, respectively.

Meanwhile, it is assumed that the $L$- and $R$-tubes have the same length $l_L = l_R = 200$ mm, but differing inner radii, $a_L = 5$ mm and $a_R = 7.5$ mm, respectively. Water at room temperature, with density $\rho = 1000$ kg/mm³ and viscosity $\mu = 0.00112$ m²/sec, is taken to be the working fluid in the system. Using then the Poiseulle's law, the flow resistance coefficients $R_{L,R}$ appearing in Eqs. (5a) and (5b) are related to the corresponding tube lengths $l_{L,R}$ and radii $a_{L,R}$ by $R_{L,R} = 8\mu l_{L,R} / \pi a_{L,R}^4$, yielding $R_L = 0.913$ Pa·sec/ml and $R_R = 0.180$ Pa·sec/ml, respectively. Moreover, the inertia coefficients $I_{L,R}$ in Eqs. (5a) and (5b) are taken to be proportional to the mass of the fluid in the $L$- and $R$-tubes: $I_{L,R} = \rho l_{L,R} / \pi a_{L,R}^2$. It is then calculated that $I_L = 2.55$ Pa·sec²/ml and $I_R = 1.13$ Pa·sec²/ml. With these parameter values, Yang and Wang (2010) identified that the dimensionless compliance parameter $\varepsilon^2 = C_{01} R_L R_R / T_{LR}(R_L + R_R)$ $= 1.03 \times 10^{-5} \ll 1$, and therefore were able to carry out some asymptotic analyses. Note also that, as in Yang and Wang (2010), here the maximum value of $V_a(t)$ is taken to be $V_{max} = 0.6 \, V_{0f}$. (The numerical results with other choices of $V_{max}$ would appear to be qualitatively similar to that discussed in the next section, and therefore will not be presented here.) Meanwhile, to avoid degenerate limiting cases, the minimum value of $V_a(t)$, $V_{min}$, is allowed to vary between $0.05 \, V_{max}$ and $0.95 \, V_{max}$ only.

## RESULTS AND DISCUSSION

### Classification of system response

Let us now discuss the numerical results of our impact dynamics model for a model valveless pumping system having the parameter values specified above. First, Fig. 3 shows the parameter boundaries for various types of system responses observed in our computations. Note that, to bring our some dynamical similarities, the numerically calculated critical frequencies are normalized by $\varepsilon^{-1}(T_L T_R)^{-1/2} = 11.83 \ \text{sec}^{-1}$ (while their dimensional values also are shown in Fig. 3 to give some ideas of their actual magnitudes). Basically, in the parameter range shown in Fig. 3, when the driving frequency $\omega$ of the actuator does not exceed the "stability-margin frequency" $\omega_{st}$, five different types of system responses are observed, namely the "pure contact" (labeled by abbreviation PC in Fig. 3), "contact–free" (CF), "purely free" (PF), "contact–free with $V_0$ reservoir depletion" (CF0), and "purely free with $V_0$ reservoir depletion" (PF0) responses.

The PC responses exist when $\omega$ is below the threshold frequency ($\omega_{th}$), in which the $V_0$ reservoir never separates from the actuator and no mean flowrate is produced in the fluid loop. When the volume ratio $V_{min}/V_{max}$ is larger, the CF responses arise when $\omega$ exceeds $\omega_{th}$ but is lower than the onset frequency $\omega_{PF}$ of the PF responses. In the CF responses, the system is in the contact mode for a certain portion of the compression cycle, and is in the free mode for the rest of the compression cycle; see Fig. 4(a) for an example. Meanwhile, in the PF responses, the $V_0$ reservoir collides with the actuator at a particular phase of the compression cycle and separates from the actuator right afterwards; see Fig. 4(b). If $\omega$ is further increased beyond the stability-margin frequency $\omega_{st}$, the system responses then become much more complicated, and will be discussed later.

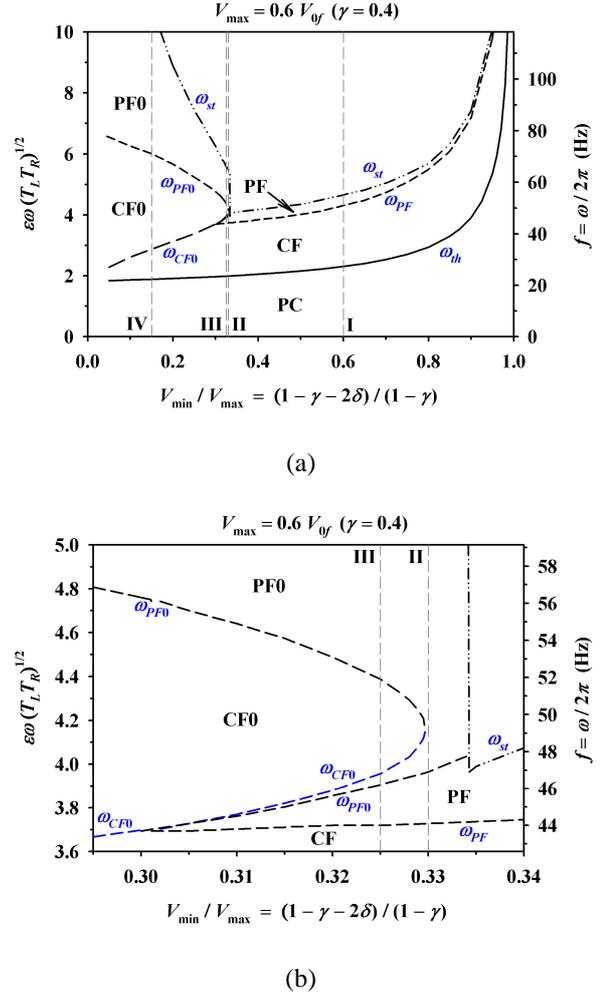

(a)

(b)

Fig. 3. Classification of system responses: (a) in the full range of $V_{min}/V_{max} \in [0.05, 0.95]$, (b) close-up in the range of $V_{min}/V_{max} \in [0.29, 0.34]$. The abbreviations for various types of system responses and the critical frequencies determining the parameter boundaries of such responses are detailed in the text.

For smaller values of the $V_{min}/V_{max}$ ratio, when $\omega$ exceeds $\omega_{th}$, the $V_0$ reservoir also would separate from the actuator during some portion of the compression cycle, so that the CF responses would arise. However, beyond a certain critical driving frequency $\omega_{CF0}$, the $V_0$ reservoir would be completely depleted at a certain phase in the free-mode portion of the CF responses. Such responses (with complete $V_0$ reservoir depletion) are referred to as the CF0 responses; see Fig. 4(c).

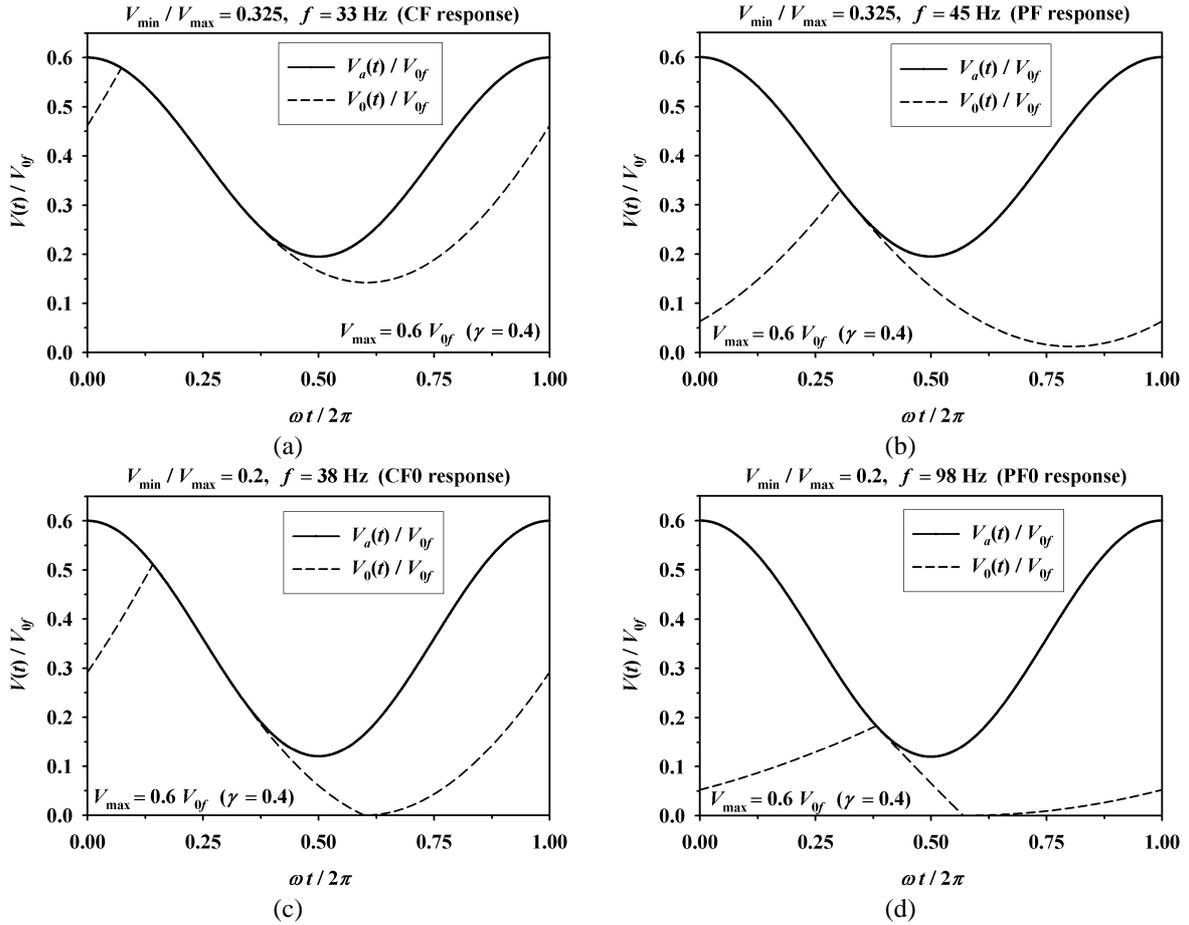

Fig. 4. Examples of various system responses. (a) The CF response, with volume ratio $V_{min}/V_{max} = 0.325$ and driving frequency $f = \omega/2\pi = 33$ Hz; (b) the PF response, with $V_{min}/V_{max} = 0.325$ and $f = 45$ Hz; (c) the CF0 response, with $V_{min}/V_{max} = 0.2$ and $f = 38$ Hz; (d) the PF0 response, with $V_{min}/V_{max} = 0.2$ and $f = 98$ Hz. Here $V_{max} = 0.6V_{0f}$ (corresponding to a compression ratio of $\gamma = 0.4$).

Moreover, when $\omega$ exceeds a even higher critical frequency $\omega_{PF0}$, the system response will become purely-free with complete depletion of the $V_0$ reservoir at a certain phase of the compression cycle. Such responses are referered to as the PF0 responses; see Fig. 4(d). And again, when $\omega$ is further increased beyond the stability-margin frequency $\omega_{st}$, the system responses become more complicated (to be discussed later).

It is also important to note that, for intermediate values of the $V_{min}/V_{max}$ ratio, the transition of system responses are somewhat more complicated. Specifically, as shown in Fig. 3(b), when $V_{min}/V_{max}$ has a value between about 0.3297 and 0.3342, the PF responses would turn into PF0 responses when the driving frequency $\omega$ of the system is increased beyond the critical frequency $\omega_{PF0}$. Meanwhile, when $V_{min}/V_{max} \in [0.3011, 0.3297]$, the PF responses turn into the PF0 responses first (when $\omega$ exceeds $\omega_{PF0}$), which, in turn, would become the CF0 responses at $\omega = \omega_{CF0}$. As the driving frequency of the system is further increased, the CF0 responses become the PF0 responses, which then lose their stability at $\omega = \omega_{st}$. In order to clarify the interrelations of the various types of system responses, in the following subsection we shall pick four particular values of $V_{min}/V_{max}$, namely 0.6, 0.33, 0.325, and 0.15 (marked by the Roman numerals I, II, III, and IV, respectively, in Fig. 3), and carefully examine the frequency dependences of a few characteristic phases of the responses (such as the particular phase at which the $V_0$ reservoir separates from or collides with the compression actuator).

Before we do that, however, an additional feature of Fig. 3(b) also is worth mentioning. Specifically, while the PF responses would turn into the PF0 responses at $\omega = \omega_{PF0}$ when the volume ratio $V_{min}/V_{max}$ is slightly less than 0.3342, for $V_{min}/V_{max} > 0.3342$ the PF responses would become seemingly chaotic at the stability-margin margin frequency $\omega_{st}$. Also, since the system essentially is restarted at the instant when the $V_0$ reservoir is completely depleted (i.e., when $V_0(t) = 0$), and therefore intuitively would be somewhat more stable, the stability-margin frequency for $V_{min}/V_{max} < 0.3342$ is much higher than that for $V_{min}/V_{max} > 0.3342$; see Fig. 3(a).

**Frequency dependences of various characteristic phases**

To begin with, let us consider the case of $V_{min}/V_{max} = 0.6$. As shown in Fig. 3(a), when the driving frequency $\omega$ increases, the PC responses would first become the CF responses at $\omega = \omega_{th}$, which, in turn, would become the PF responses at $\omega = \omega_{PF}$. Also, when $\omega$ exceeds $\omega_{st}$, more complicated system responses would be excited. So, for $\omega \in (\omega_{th}, \omega_{st})$, there are two important characteristic phases of the system responses, namely the phase $\theta_s$ ($\in [0, 2\pi]$) of the actuator motion at which it separates from the $V_0$ reservoir, and the phase $\theta_c$ at which the actuator collides with the $V_0$ reservoir. To examine the driving-frequency dependences of such characteristic phases, and bring out their interrelations at the same time, we find it convenient to use polar plots.

Specifically, in Fig. 5(a), the radial variable is the excess of the driving frequency $f = \omega/2\pi$ relative to the threshold frequency $f_{th} = \omega_{th}/2\pi$, while the angular variable is simply the phase of the actuator motion. It is clearly seen in Fig. 5(a) that the $\theta_s$ and $\theta_c$ curves both start at the threshold frequency $f = f_{th}$, with the common value of about $180°$. This, in fact, is easy to understand, because for $f < f_{th}$ the $V_0$ reservoir would never separate from the actuator (and therefore be impossible to collide with it). As the driving frequency (i.e., the radial variable) increases, the value of $\theta_s$ decreases, while that of $\theta_c$ increases. Accordingly, the actual duty cycle (i.e., the fraction of the compression cycle in which the actuator is in solid contact with the $V_0$ reservoir) decreases with increasing $f$. Then, at $f = f_{PF} = \omega_{PF}/2\pi$ (corresponding to $f - f_{th} = 23.3$ Hz in Fig. 5(a)), the actual duty cycle becomes zero, in which case the $V_0$ reservoir collides with the actuator at a certain phase of the compression cycle and rebounds from the actuator right afterwards. So, logically, in such cases the separation $\theta_s$ is also the collision phase $\theta_c$. As a result, for $f \in (f_{th}, f_{st})$ (with $f_{st} = \omega_{st}/2\pi$) it suffices to use a single "rebound phase" ($\theta_r$) to characterize the system response. Note also that when the driving frequency $f$ exceeds the stability-margin frequency $f_{st} = \omega_{st}/2\pi$, the system responses would not repeat themselves in one compression cycle, and would appear to be chaotic. Such responses will be discussed later.

Consider now the case of $V_{min}/V_{max} = 0.33$. As shown in Figs. 3(a) and 3(b), for driving frequencies $f < f_{PF}$ (with $f_{PF} = 41.1$ Hz, or $f_{PF} - f_{th} = 20.5$ Hz) the transitions of the system responses are similar to that in the previous case (with $V_{min}/V_{max} = 0.6$). So, in the characteristic phase plot, Fig. 5(b), the evolution of the $\theta_s$, $\theta_c$, and $\theta_r$ curves with the driving frequency for $f < f_{PF}$ is similar to that in Fig. 5(a). However, in the present case, when $f$ exceeds $f_{PF}$, the system responses would first experience a complete depletion of the $V_0$ reservoir (at $f = f_{PF0} = \omega_{PF0}/2\pi$) before they become seemingly chaotic when $f > f_{st}$. Accordingly, in Fig. 5(b), the curve for the characteristic phase $\theta_0$ at which the $V_0$ reservoir is completely depleted appears at $f = 46.9$ Hz (or $f - f_{th} = 23.3$ Hz). It is also interesting to note that, since the system responses undergo a qualitative change (from PF to PF0) at such a critical frequency, the $\theta_r$ curve then makes a sharply turn there; see Fig. 5(b). Meanwhile, since the system essentially is restarted after the $V_0$ reservoir is completely depleted, the stability-margin frequency $f_{st}$ is significantly increased (compared with that in the previous case of $V_{min}/V_{max} = 0.6$). Still, the PF0 responses would become seemingly chaotic (which will be discussed later) when $f$ exceeds $f_{st}$.

Next, the characteristic phases for the case of $V_{min}/V_{max} = 0.325$ are plotted against the driving frequency $f$ in Fig. 5(c). To understand the evolution of the curves in Fig. 5(c), let us observe first in Figs. 3(a) and 3(b) that for $\omega > \omega_{PF0}$ (the lower branch), the system responses would transit from PF0 to CF0 (at $\omega = \omega_{CF0}$), and then from CF0 back to PF0 (at $\omega = \omega_{PF0}$, the upper branch), before they eventually become seemingly chaotic at $f = f_{st}$. Accordingly, in Fig. 5(c), the $\theta_r$ curve first bifurcates into a $\theta_s$ curve and a $\theta_c$ curve at $f = f_{PF0} = \omega_{PF0}/2\pi = 47.0$ Hz (or, $f - f_{th} = 23.4$ Hz), which then merge into a single $\theta_r$ curve at $f = f_{PF0} = 51.9$ Hz (or, $f - f_{th} = 28.3$ Hz). And again, the PF0 responses would eventually become seemingly chaotic when $f$ exceeds $f_{st}$.

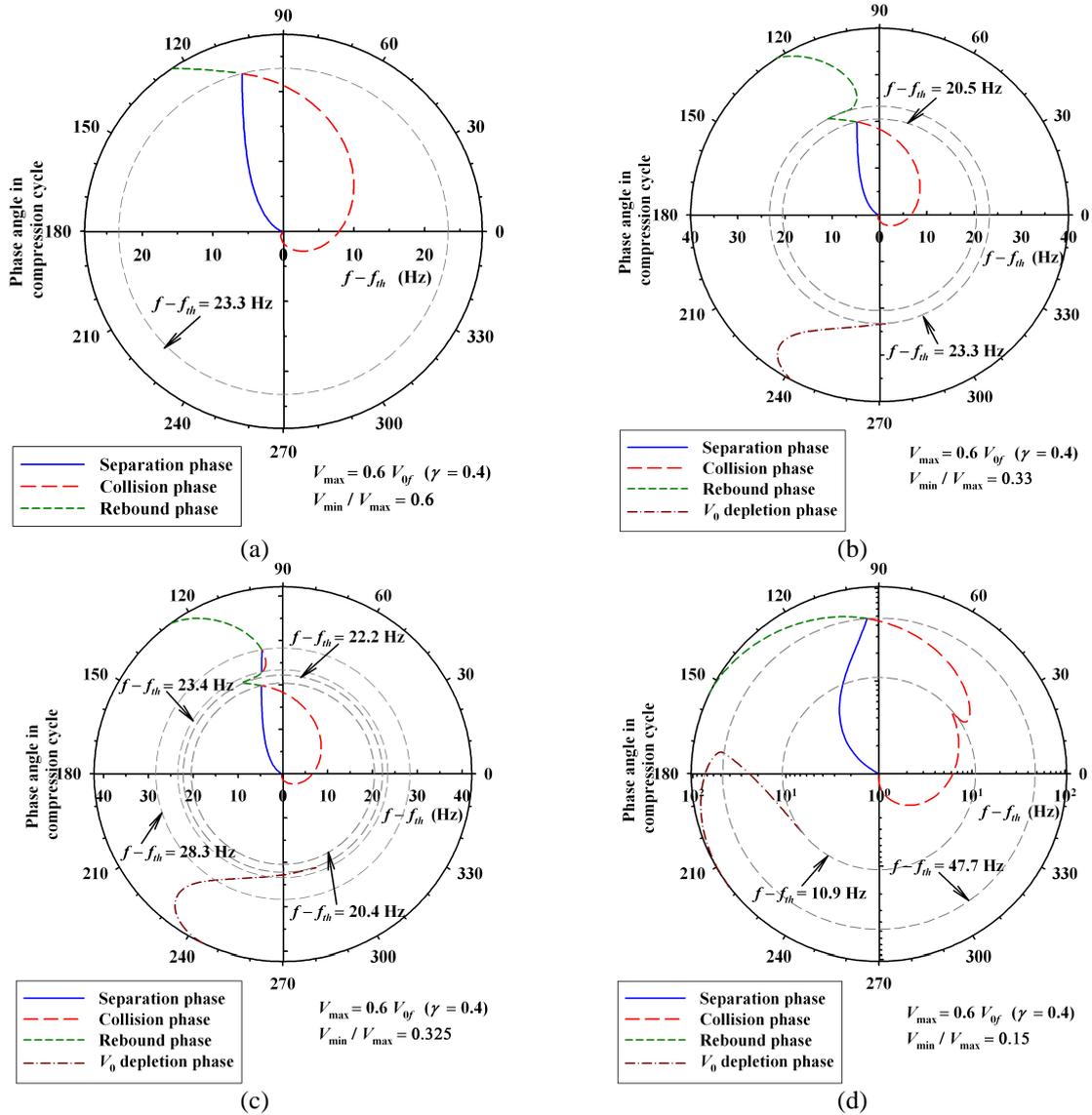

Fig. 5.  Frequency dependences of various characteristic phases of the system responses for driving frequencies $f \in (f_{th}, f_{st})$. (a) Case I: $V_{min}/V_{max} = 0.6$, with $f_{th} = 27.32$ Hz and $f_{st} = 55.52$ Hz; (b) Case II: $V_{min}/V_{max} = 0.33$, with $f_{th} = 23.63$ Hz and $f_{st} = 64.33$ Hz; (c) Case III: $V_{min}/V_{max} = 0.325$, with $f_{th} = 23.59$ Hz and $f_{st} = 65.89$ Hz; (d) Case IV: $V_{min}/V_{max} = 0.15$, with $f_{th} = 22.35$ Hz and $f_{st} = 127.85$ Hz.

Finally, for the case of $V_{min}/V_{max} = 0.15$, the characteristic phases are plotted against the driving frequency $f$ in Fig. 5(d).  (Note that, as the range of the driving frequency $f$ is now much larger than that in previous cases, here the value of $f$ is plotted in logarithmic scales.)  Compared with the previous case (with $V_{min}/V_{max} = 0.325$), here the transition of the system responses are much simpler, namely from PC to CF (at $f = f_{th}$), CF0 (at $f = f_{CF0}$), PF0 (at $f = f_{PF0}$) in turn, and finally to the seemingly chaotic responses (at $f = f_{st}$).  Also, since all the important features of the characteristic phase curves have been explained above, they are readily understandable in Fig. 5(d), and therefore will not be over-discussed here.

**Seemingly chaotic system responses**

We now briefly discuss some interesting features of the seemingly chaotic responses of the system when the driving frequency $f$ exceeds the stability-margin frequency $f_{st}$.  Note that, as such responses seem to be chaotic, they are quite sensitive to the specific parameter values of the system.  However, for the purpose of demonstrating a number of qualitative features of such responses, here we shall only consider the case of $V_{min} / V_{max} = 0.33$. It is also important to note that, for such responses the temporal means (over one compression cycle) of the flowrates in the system would vary from cycle to cycle. So, to extract some orders of the seemingly chaotic responses, for each parameter setting the numerical computations are run for 2000 compression cycles, and the temporal means of the aforementioned system state variables are computed for each compression cycle. Of course, during the first few cycles, the variations in such temporal means may result partly from the initial transients. However, we found that the initial transient effects practically are negligible after a few dozen cycles. Also, it suffices to keep track of the last 50 cycles (of the 2000 cycles of computation) to bring out the general pictures of the system responses.

To illustrate the frequency dependences of the system responses for $f > f_{st}$, it remains useful to trace the variations of the critical phase angles $\theta_s$, $\theta_c$, $\theta_r$, and $\theta_0$, as shown in Fig. 6(a). Note, however, that such phases generally vary from cycle to cycle, and that the results for the last 50 cycles are plotted. So, for a given driving frequency $f$ (which can be represented by a circle in Fig. 6(a)), a certain critical phase (say $\theta_r$) may have multiple values and hence be represented by several data points. Moreover, for $f > f_{st}$, the critical phases are extremely sensitive to a small variation in $f$. As a result, the data points in Fig. 6(a) appear to be scattering, and generally do not form continuous curves (while they do in Figs. 5(a)–5(d)). (Note also that here the computations are carried out at the driving-frequency increment of 1 Hz.)

However, some characteristic phases do seem to form continuous curves within a number of driving-frequency windows. For example, in the frequency window bounded approximately by 109 Hz and 177 Hz (labeled with "$2T$" in Fig. 6(a)), the separation phase ($\theta_s$) and collision phase ($\theta_c$) appear to form two continuous curves, which start at about $f = 109$ Hz and merge into a single curve of rebound phase ($\theta_r$) at about $f = 135$ Hz. Meanwhile, the zero volume phase ($\theta_0$) also appears to form a continuous curve throughout this frequency window. Moreover, in the frequency window bounded approximately by 226 Hz and 266 Hz (labeled with "$3T$"), the rebound phase ($\theta_r$) and zero volume phase ($\theta_0$) appear to form two continuous curves. Some insights about the dynamical behaviors of the "$2T$" and "$3T$" responses can be gained by plotting the mean flowrate in the $L$-tube, $\langle Q_L \rangle$ (normalized by the characteristic flowrate scale $\Delta V_a / T_{LR}$), against the driving frequency (subtracted by the threshold frequency $f_{th}$); see Fig. 6(b). Two examples of such responses are indicated by lines (a) and (b) in Fig. 6(b). As can be readily seen, for the "$2T$" and "$3T$" responses, the mean flowrates of the 50 cycles plotted in Fig. 6(b) appear to repeat every two cycles and every three cycles, respectively, and that is exactly why the responses are named as such. In fact, while being somewhat more difficult to observe in Fig. 6(a), some more regularities in the system responses can be observed in Fig. 6(b); two examples are indicated by lines (c) and (d) there. Specifically, on lines (c) and (d), the mean flowrates of the 50 plotted cycles appear to repeat every three cycles and every five cycles, respectively. The regularities of such responses also can be seen more clearly by plotting them in time domain; see Figs. 7(a) –(d).

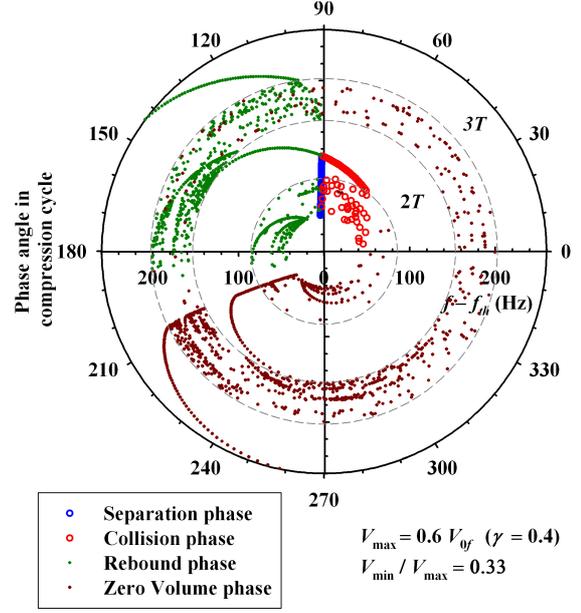

(a)

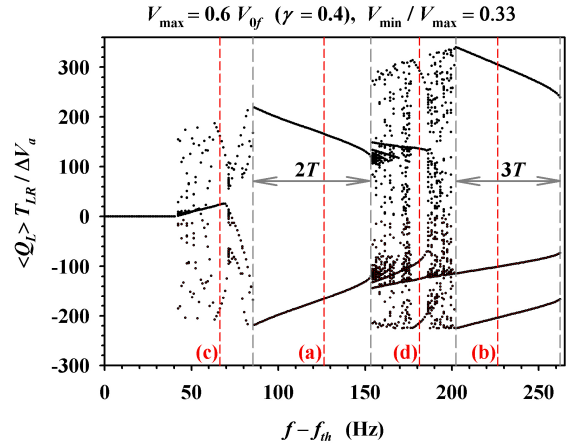

(b)

Fig. 6. Frequency dependences of various characteristic phases. (a) and mean flow rate in the $L$-tube (b) for system responses with driving frequencies $f > f_{st}$. Here $V_{min} / V_{max} = 0.33$, with $f_{th} = 23.63$ Hz and $f_{st} = 64.33$ Hz.

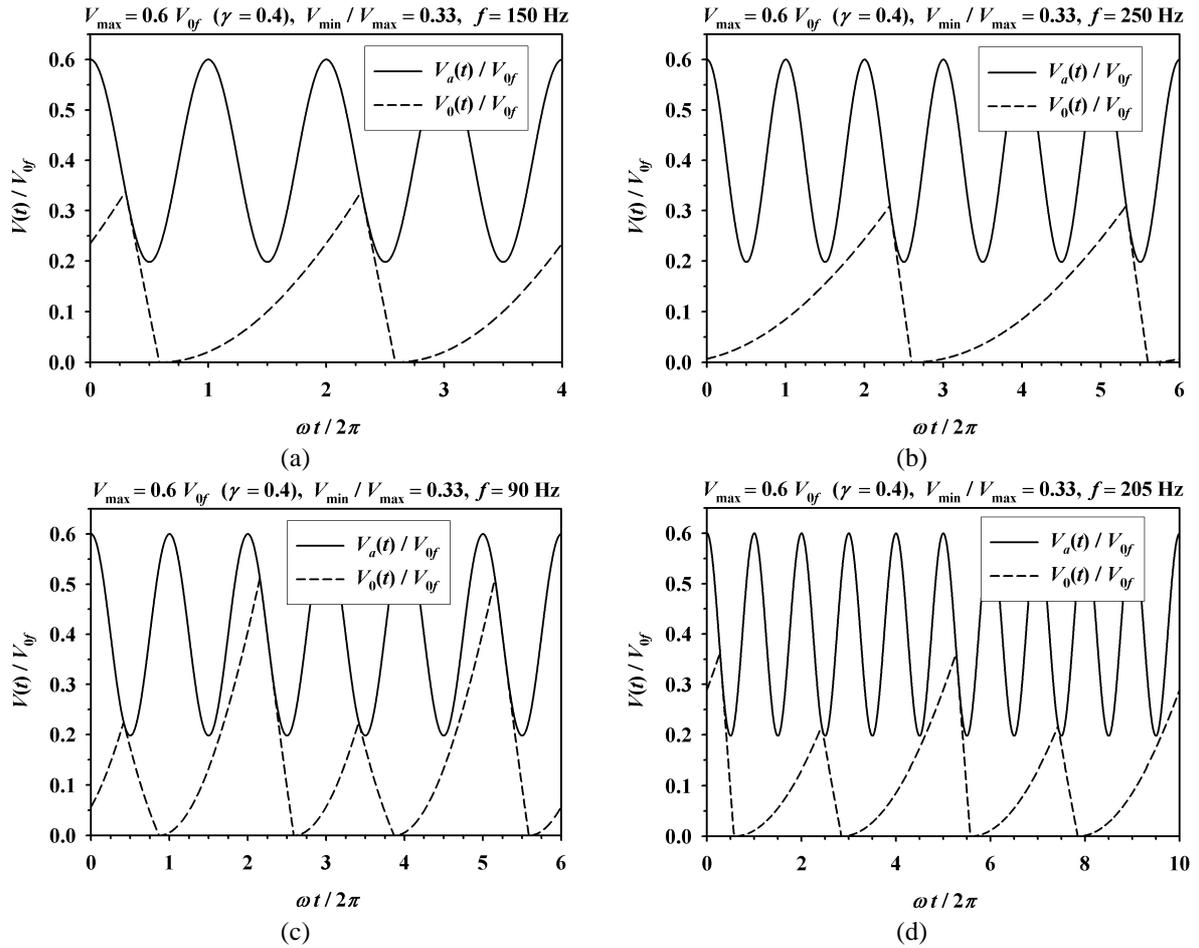

Fig. 7. Various system responses for $V_{min}/V_{max} = 0.33$ with driving frequencies of (a) 150 Hz, (b) 250 Hz, (c) 90 Hz, and (d) 205 Hz, respectively.

Of course, there certainly are other less obvious regularities of the system responses in the frequency range plotted in Figs. 6(a) and 6(b), and much much more at higher driving frequencies. Meanwhile, outside the particular frequency windows within which such regular responses exist, the seemingly chaotic responses can get as complicated as one can imagine. However, as a detailed analysis of such response clearly is non-trivial, and is beyond the scope of this work, it will not be further pursued. Instead, let us point out an additional observation in Fig. 6(a). Specifically, it is seen in Fig. 6(a) that the $V_0$ reservoir may stay in contact with the compression actuator only when the driving frequency is lower than about 135 Hz. Above that frequency, the separation–collision interaction between the $V_0$ reservoir and the actuator would no longer be possible (and the separation and collision phases cease to appear in Fig. 6 (a)). What happen then are that, whenever the $V_0$ reservoir collides with the actuator they separate from each other immediately afterwards, and that each collision of them would be so strong to completely deplete the $V_0$ reservoir. As a result, in Fig. 6(a), for $f > 135$ Hz, only the rebound and zero volume phases exist and are plotted there.

Finally, to conclude this discussion, let us speculate a few possible applications of the system responses with driving frequencies $f > f_{st}$. First, as can be readily seen in Fig. 6(b), the mean flowrates (in each cycle) for $f > f_{st}$ are greater than that for $f < f_{st}$ by several orders of magnitude. In fact, the difference is so huge that the mean system flowrates for $f < f_{st}$ cannot be discerned in the linear plot of Fig. 6(b). But the results for $f < f_{st}$ have been plotted in logarithmic scale and carefully discussed in Yang and Wang (2010); the interested reader is referred to that work for further details. Note, however, that since the mean flowrates for $f > f_{st}$ vary from cycle to cycle, and that their overall (i.e., long-term) averages actually have the same order of magnitude as that for $f < f_{st}$. This finding may have an important engineering bearing. Specifically, if one would like to design a micro-pump with a stable flowrate output, it is advisable to choose a driving frequency $f$ that is lower than the stability-margin frequency $f_{st}$, because increasing $f$ would likely induce the seemingly chaotic system responses but would not increase the long-term average of the system flowrate. However, if one wishes to design a micro-mixer, the large yet irregular flowrate oscillation for $f > f_{st}$ may be useful for enhancing the mixing effects. It is important to note, however, that some further works still are needed to assess the feasibility of such a possibility.

## CONCLUDING REMARKS

In this work, using the lumped-parameter model of Yang and Wang (2010) for a valveless pumping system with actuator impact effects, we have carried out a more comprehensive numerical study than that discussed in Yang and Wang (2010). For periodic system responses repeating themselves in one compression cycle, five different types of responses are found, which are referred to above as the pure contact (PC), contact–free (CF), purely free (PF), contact–free with $V_0$ reservoir depletion (CF0), and purely free with $V_0$ reservoir depletion (PF0) responses (and illustrated in Fig. 4). By examining the dependences on the volume ratio $V_{min}/V_{max}$ of a few critical frequencies at which the system responses undergo a transition in type, the parameter boundaries within which such responses exist have been clarified (see Fig. 3). Moreover, for a few particular values of $V_{min}/V_{max}$, the driving-frequency dependences of a number of characteristic phases, namely the separation phase $\theta_s$, collision phase $\theta_c$, rebound phase $\theta_r$, and $V_0$ reservoir depletion phase $\theta_0$ (see Fig. 5), the interrelations of the different types of system responses have been identified. It is believed that these results not only are useful for understanding the nonlinear impact dynamics of valveless pumping systems, but also would help optimize the design of practical valveless pumps with certain prescribed specifications.

It has also been demonstrated in this work that, for driving frequencies $f$ higher than the stability-margin frequency $f_{st}$, the system responses generally do not repeat themselves after one compression cycle. Instead, they appear to be "chaotic." For such seemingly chaotic system responses, several interesting features have been observed. For example, we have found several driving-frequency windows within which the system responses appear to repeat themselves after an integral multiple of the compression period (see Figs. 6 and 7). Moreover, the mean flowrates (in each cycle) for $f > f_{st}$ generally are greater than that for $f < f_{st}$ by several orders of magnitude, while their overall (i.e., long-term) average has the same order of magnitude as that for $f < f_{st}$. This finding suggests that, if one would like to design a valveless pump (say, for certain microfluidic or biomedical applications) with a stable flowrate output, it would be advisable to choose a driving frequency $f$ that is lower than the stability-margin frequency $f_{st}$. However, if one wishes to design a micro-mixer instead, the large yet irregular flowrate oscillation for $f > f_{st}$ may be useful for enhancing the mixing effects. And of course, some further works still are needed to assess the feasibility of such possibilities.

Finally, to conclude this paper, we would also like to point out that the impact dynamics model used here (and in Yang and Wang, 2010) has suppressed many other mechanisms that may produce significant valveless pumping effects (such as that of impedance pumps; see, for example, Hickerson and Gharib, 2006; Hickerson et al., 2005; Huang et al, 2010; Wen and Chang, 2009). The model has been designed in this particular way to demonstrate the complicated actuator impact dynamics of valveless pumping systems with the highest possible clarity. Given the many interesting findings discussed above, in future work it certainly is worth pursuing to design a suitable experimental apparatus for their verification. Meanwhile, in many realistic microfluidic or biomedical applications, one may also need to take into account the synergetic interactions of the actuator impact effect with other valveless pumping mechanisms, which is also worth pursuing in future work.

## ACKNOWLEDGEMENT

The authors gratefully acknowledge the Taiwan National Science Council for supporting this work through grant NSC96-2628-E-006-241-MY3. They would also like to thank Professors M.-S. Ju, S.-Y. Lee, K.-S. Chen, and C.-Y. Wen of NCKU for a number of fruitful discussions on this work and other related topics.

# NOMENCLATURE

| | |
|---|---|
| $a_{L,R}$ | radii of the $L$- and $R$-tubes |
| $C_{0,1}$ | compliance coefficients of reservoirs $V_0$ and $V_1$ |
| $C_{01}$ | characteristic compliance of the closed loop |
| $f$ | driving frequency of the actuator |
| $f_{st}$ | stability-margin frequency of steady periodic responses |
| $f_{th}$ | threshold frequency for actuator separation |
| $h_{0,1}$ | free heights of reservoirs $V_0$ and $V_1$ |
| $I_{L,R}$ | inertia coefficients of the $L$- and $R$-tubes |
| $k_{0,1}$ | effective spring constants of reservoirs $V_0$ and $V_1$ |
| $l_{L,R}$ | lengths of the $L$- and $R$-tubes |
| $p_{0,1}(t)$ | fluid pressures inside reservoirs $V_0$ and $V_1$ |
| $p_e(t)$ | external pressure of reservoir $V_0$ |
| $Q_{L,R}(t)$ | volumetric flowrates in the $L$- and $R$-tubes |
| $R_{L,R}$ | resistance coefficients of the $L$- and $R$-tubes |
| $r_{0,1}$ | radii of reservoirs $V_0$ and $V_1$ |
| $T$ | period of the actuator motion |
| $t$ | time |
| $T_{L,R}$ | intrinsic time scales of the $L$- and $R$-tubes |
| $T_{LR}$ | characteristic time scale of the closed loop |
| $V_a(t)$ | admissible $V_0$ reservoir volume set by the actuator motion |
| $V_{\max}, V_{\min}$ | maximum and minimum values of $V_a(t)$ |
| $V_{0,1}(t)$ | instantaneous volumes of reservoirs $V_0$ and $V_1$ |
| $V_{0f}, V_{1f}$ | free volumes of reservoirs $V_0$ and $V_1$ |
| $\gamma$ | pre-compression ratio |
| $\Delta V_a$ | half-stroke volume of the actuator motion |
| $\varepsilon^2$ | dimensionless compliance parameter |
| $\mu$ | viscosity of working fluid (water) |
| $\rho$ | density of working fluid (water) |

# 無閥流體迴路在擠壓致動器衝擊作用下之動態響應

王齊中　　楊天祥
國立成功大學機械工程學系

## 摘　要

無閥流體驅動現象廣泛存在於工程系統中與生物體內,且為其中工作流體輸送之重要助力、甚或主力。在先前研究中,我們曾以一由兩條硬管連結兩個可延展流體儲存槽所組成之封閉無閥流體迴路為模擬標的,建構了一套分段線性之總括參數數學模型,並用以釐清擠壓致動器衝擊效應對無閥流體驅動系統動態表現的影響。該研究所獲得之漸近分析結果與初步數值計算結果指出,擠壓頻率以及其他系統參數對於致動器與受壓槽間的交互作用具有重要的影響,進而使得系統動行為甚為豐富多樣。在本文中我們針對前述數學模型進行更為完整的數值參數探討,並且有系統地討論計算結果,從而歸納出系統動態響應之可能類型,以及各類型響應所對應之系統參數範圍。同時,藉由追蹤若干特徵相位角(例如在一擠壓週期中,致動器與受壓槽脫離或碰撞時的相位角)隨擠壓頻率變動之趨勢,我們也明確地指認各類型系統動態響應的關連性與演進過程。